\documentclass[twocolumn,showpacs,preprintnumbers,amsmath,amssymb,pre]{revtex4}

\usepackage{epsfig}
\usepackage{bm}

\newcommand{\EF}{\epsilon_F}

\newcommand{\nion}{n_\mathrm{ion}}
\newcommand{\Nion}{N_\mathrm{ion}}
\newcommand{\kB}{k_B}

\newcommand{\omp}{\omega_p}

\newcommand{\pF}{p_F}
\newcommand{\TF}{T_F}

\newcommand{\Tm}{T_m}
\newcommand{\Tp}{T_p}
\newcommand{\Tr}{T_r}
\newcommand{\xr}{x_\mathrm{rel}}
\newcommand{\gr}{\gamma_r}

\newcommand{\beq}{\begin{equation}}
\newcommand{\eeq}{\end{equation}}
\newcommand{\bea}{\begin{eqnarray}}
\newcommand{\eea}{\end{eqnarray}}
\newcommand{\req}[1]{Eq.\ (\ref{#1})}

\newcommand{\ApJ}[1]{{Astrophys.\ J.} \textbf{#1}}
\newcommand{\ApJS}[1]{{Astrophys.\ J. Suppl.\ Ser.} \textbf{#1}}

\newcommand{\PR}[2]{Phys.\ Rev. #1 \textbf{#2}}

\begin{document}


\title{Equation of state of classical Coulomb plasma mixtures}

\author{Alexander Y. Potekhin}
 \altaffiliation[Also at the ]{Ioffe Physical-Technical Institute,
     194021 St.\ Petersburg, Russia}
     \email{palex@astro.ioffe.ru}
\affiliation{Ecole Normale Sup\'erieure de Lyon,
     CRAL (UMR CNRS No.\ 5574),
     69364 Lyon Cedex 07, France}

\author{Gilles Chabrier}\email{chabrier@ens-lyon.fr}
    \affiliation{Ecole Normale Sup\'erieure de Lyon,
     CRAL (UMR CNRS No.\ 5574),
     69364 Lyon Cedex 07, France}
 
\author{Forrest J. Rogers}
\affiliation{Lawrence Livermore National Laboratory,
 P. O. Box 808, Livermore, CA 94550, USA}

\date{\today}


\begin{abstract}
We develop analytic approximations of thermodynamic  functions of
fully ionized nonideal electron-ion plasma mixtures. In the
regime of strong Coulomb coupling, we use our previously
developed analytic approximations for the free energy of
one-component plasmas with rigid and polarizable electron
background and apply the linear mixing rule (LMR). Other
thermodynamic functions are obtained through analytic derivation
of this free energy. In order to obtain an analytic approximation
for the intermediate coupling and transition to the
Debye-H\"uckel limit, we perform hypernetted-chain calculations
of the free energy, internal energy, and pressure for  mixtures
of different ion species and introduce a correction to the LMR,
which allows a smooth transition from strong to weak Coulomb
coupling in agreement with the numerical results. 

\end{abstract}



\maketitle

\section{Introduction}
\label{sect:intro}

We study the equation of state (EOS) of fully ionized nonideal
electron-ion plasmas (EIP). In a previous work \cite{CP98,PC00}, hypernetted
chain (HNC) calculations were performed and analytic formulae
were proposed for EOS calculations of EIP containing a single ion species.
For \emph{mixtures} of different ion species, the 
EOS was calculated using the linear mixing rule
(LMR), whose high accuracy at $\Gamma>1$
was previously confirmed in a number of studies
\cite{HTV77,ChabAsh,Rosenfeld,DWSC96,DWS03}. However, the LMR is
inaccurate for weakly coupled plasmas. Some consequences of its
violation were studied by Nadyozhin and Yudin
\cite{NadyozhinYudin05}, who showed that the  differences between
the linear and nonlinear mixing at moderate Coulomb coupling
($0.1\lesssim\Gamma\lesssim1$) can shift the nuclear
statistical equilibrium at the final stage of a stellar
gravitational collapse.

In this paper, we perform HNC
calculations of the free energy, internal energy, and pressure
for mixtures of various kinds of ions in the weak, intermediate,
and strong coupling regimes, and suggest an analytic correction
to the LMR.

In Sec.~\ref{sect:plasmpar} we define the basic plasma parameters. 
In Sec.~\ref{sect:mix}
we calculate the EOS of ion mixtures and propose an analytic
formula for the EOS of multicomponent EIP, applicable at any
$\Gamma$ values. The summary is given in Sec.~\ref{sect:concl}.

\section{Plasma parameters}
\label{sect:plasmpar}

Let $n_e$ be the electron number density and
$n_j$ the number density of ion species $j$=1,2,\ldots,
with mass and charge numbers $A_j$ and $Z_j$, respectively.
The total number density of ions is $\nion = \sum_j n_j$.
The electric neutrality implies $n_e = \langle Z \rangle \nion$.
Here and hereafter the brackets $\langle \ldots \rangle$ denote
averaging of the type
$
   \langle f \rangle = \sum_j x_j f_j ,
$
where $x_j\equiv n_j/\nion$.

The state of a free electron gas is determined by the electron
number density $n_e$ and temperature $T$. Instead of $n_e$ it is
convenient to introduce the dimensionless density parameter
$r_s=a_e/a_0$, where $a_e=(\frac43\pi n_e)^{-1/3}$
and $a_0$ is the Bohr radius.

At stellar densities it is convenient to use, instead of $r_s$,
the relativity parameter \cite{Salpeter61}
$\xr  = \pF / m_e c = 0.014\,r_s^{-1}$,
where $ \pF = \hbar \,(3 \pi^2 n_e)^{1/3}$
is the electron Fermi momentum.
The Fermi kinetic
energy is
$\EF = c\,\sqrt{(m_e c)^2 + \pF^2}-m_e c^2 ,
$
and the Fermi temperature equals
$\TF \equiv \EF/\kB = \Tr  \, ( \gr - 1) ,$
where 
$\Tr \equiv {m_e c^2 / \kB } = 5.93\times
10^9~\mathrm{K}$,
$\gr \equiv \sqrt{1+ \xr^2}$,
and $\kB$ is the Boltzmann constant.

The ions are nonrelativistic in most
applications. The strength of the Coulomb interaction of ion
species $j$ is characterized by the Coulomb coupling parameter,
\begin{equation}
   \Gamma_j =
        (Z_j e)^2/(a_j  \kB T) = \Gamma_e Z_j^{5/3},
\end{equation}
where
$a_j  = a_e Z_j^{1/3}$
is the ion sphere radius and 
$
    \Gamma_e \equiv { e^2 }/({ a_e \kB T}).
$
In a multicomponent plasma,
it is useful to introduce the mean ion-coupling parameter
$ \Gamma = \Gamma_e \langle Z^{5/3} \rangle$ \cite{HTV77}.
At a \emph{melting temperature}
$\Tm$, corresponding to $\Gamma\approx175$ (e.g., \cite{PC00}),
the plasma freezes into a Coulomb crystal.

An important scale length 
is the thermal de Broglie wavelength
$
   \lambda_j= ({2\pi\hbar^2}/{ m_j \kB T})^{1/2},
$
where $m_j$ is the ion mass. 
The electron thermal length $\lambda_e$ is given by
the same expression with $m_j$ replaced by $m_e$.

The quantum effects on ion motion become
important at
$T \ll \Tp$, where
$
   \Tp \equiv {\hbar \omp }/{ \kB}
$
and
$
    \omp = \left(  {4 \pi e^2 \,\nion}
    \left\langle { Z^2 /m } \right\rangle \right)^{1/2}
$
is the ion plasma frequency. 

In this paper we consider only
the classical Coulomb liquid, which implies $T\gg\Tp$ and
$T>\Tm$.

\section{Equation of state}
\label{sect:mix}

Assuming commutativity of the kinetic
and potential operators and separation of the
traces of the electronic and ionic parts of
the Hamiltonian,
the total Helmholtz free energy $F$
can be conveniently written as
\beq
   F = 
   F_\mathrm{id}^{\mathrm{ion}} + F_\mathrm{id}^{(e)} 
   + F_{ee} + F_{ii} + F_{ie}, 
\label{f-tot}
\eeq
where $F_\mathrm{id}^{\mathrm{ion}}$ and $F_\mathrm{id}^{(e)}$  denote
the ideal free energy of ions and electrons,  and the last three
terms represent an excess free energy arising from  the
electron-electron, ion-ion, and ion-electron interactions,
respectively. 

The pressure $P$,
the internal energy $U$, and the entropy $S$ of an ensemble of
fixed number of plasma particles in volume $V$ can be
obtained using the thermodynamic relations
$
   P=-(\partial F/\partial V)_T, \quad
   S= -(\partial F / \partial T )_V,
$
and $U=F+TS$.
The second-order thermodynamic functions are derived by
differentiating these first-order ones. The
decomposition (\ref{f-tot}) induces the analogous decomposition
of $P$, $U$, $S$, the heat capacity
$C_V=(\partial S/\partial\ln T)_V,$ and
the logarithmic pressure derivatives
$
   \chi_T=(\partial\ln P/\partial\ln T)_V
$ 
and
$
   \chi_\rho=-(\partial\ln P/\partial\ln V)_T.
$
Other second-order functions can be expressed through these
by Maxwell relations (e.g., \cite{LaLi-SP1}).

\subsection{Ideal electron-ion plasmas}
\label{sect:id}

The free energy of a gas of
$N_j=n_j V$ nonrelativistic classical ions of $j$th kind is
\beq
   F_\mathrm{id}^{(j)} =
     N_j \kB T\,\left[\ln(n_j\lambda_j^3/g_j)-1 \right],
\label{id_i}
\end{equation}
where $g_j$ is the spin multiplicity. The
total free energy is given by the sum
$F_\mathrm{id}^{\mathrm{ion}}=\sum_j F_\mathrm{id}^{(j)}$.
Analogous sums give $U$, $S$,  $P$,
and $C_V$. Since \req{id_i} contains $n_j$ under
logarithm, these sums for $F$ and $S$ naturally include
the entropy of mixing  $S_\mathrm{mix}=-\kB\sum_j N_j \ln x_j$.

The free energy of the electron gas is given by
\beq
   F_\mathrm{id}^{(e)} =
   \mu_e N_e  - P_\mathrm{id}^{(e)}\,V,
\label{id_e}
\end{equation}
where $\mu_e$ is the electron chemical potential.
The pressure $P_\mathrm{id}^{(e)}$ and the number density
$n_e=N_e/V$
are functions of $\mu_e$ and $T$, which can be written through
the Fermi-Dirac integrals $I_\nu(\chi_e,\tau)$, where 
$\chi_e=\mu_e/\kB T$ and $\nu=1/2$, $3/2$, and $5/2$. In
Ref.~\cite{CP98} we gave analytic approximations for the
Fermi-Dirac integrals, based on fits \cite{Blin} to
electron-positron thermodynamic functions.
The chemical potential at a given density can be found either
from a numerical inversion of function $n_e(\chi_e,T)$ or using the
analytic approximation \cite{CP98}.
The electron-gas contributions to $\chi_T$, $C_V$, and $S$ tend
to zero at $T\ll\TF$. Related numerical problems and their cure
will be discussed elsewhere \cite{EOS3}.

\subsection{Nonideal plasmas containing one type of ions}

Let us recall fit formulae for nonideal EIP containing a
single kind of ions.

\paragraph{Electron exchange and correlation.}
Electron-electron (exchange-correlation) effects were
studied by many authors. For the reasons explained in
Refs.~\cite{CP98,EOS3}, we adopt the fit to 
$f_{ee}\equiv F_{ee}/(N_e\kB T)$ presented in
Ref.~\cite{IIT}.

\paragraph{One-component plasma.}
The internal energy of the liquid one-component plasma (OCP) at
any values of $\Gamma$ is given by \cite{PC00}
\beq
  u_{ii}
  = \Gamma^{3/2}
   \left(\frac{A_1}{\sqrt{A_2+\Gamma}}
   + \frac{A_3}{ 1+\Gamma}\right)
      + \frac{B_1\,\Gamma^2 }{ B_2+\Gamma}
      + \frac{B_3\,\Gamma^2 }{ B_4+\Gamma^2},
\label{fitionu}
\eeq
where
$u_{ii} \equiv U_{ii}/\kB T \Nion$,
and 
\beq
   A_3=-\sqrt{3}/2-A_1/\sqrt{A_2}
\label{A3}
\eeq
ensures the correct
transition to the Debye-H\"uckel limit. The parameters
$A_1=-0.907347$, $A_2=0.62849$, $B_1=0.0045$, $B_2=170$,
$B_3=-8.4\times10^{-5}$, and $B_4=0.0037$ allow one to reproduce
the best available MC simulations of liquid OCP at
$1\leq\Gamma\leq190$ \cite{Caillol} with an accuracy matching the
numerical MC noise.

From \req{fitionu} one obtains the analytic expression for
$f_{ii} \equiv F_{ii}/\kB T \Nion$ by integration, and then the
Coulomb contributions to the other thermodynamic functions by
differentiation \cite{PC00}.

\paragraph{Electron polarization.}
Electron polarization in Coulomb liquid was studied by
perturbation \cite{GalamHansen,YaSha} and HNC \cite{ChabAsh,CP98,PC00}
techniques. The results  for $f_{ie} \equiv F_{ie}/ \Nion \kB T$
have been fitted by the expression \cite{PC00}
\beq
   f_{ie} = -\Gamma_e \,
   \frac{ c_\mathrm{DH} \sqrt{\Gamma_e}+
    c_\mathrm{TF} a \Gamma_e^\nu g_1(r_s,\Gamma_e) g_3(\xr)
    }{
     1+\left[ b\,\sqrt{\Gamma_e}+ a g_2(r_s,\Gamma_e) \Gamma_e^\nu/r_s \right]
              \gr^{-1} }.
\label{fitscr}
\end{equation}
The coefficients $c_\mathrm{DH}$, $c_\mathrm{TF}$, $a$, $b$, $\nu$ and functions
$g_{1,2}(r_s,\Gamma_e)$ and $g_3(\xr)$ parametrically depend on
the ion charge $Z$. Here the
coefficients $c_\mathrm{DH}$ and $c_\mathrm{TF}$ are \emph{not}
free fit parameters, because
\beq
   c_\mathrm{DH} = (Z/\sqrt{3})\,[(1+Z)^{3/2}-1-Z^{3/2}]
\label{cDH}
\eeq
ensures the transition of the excess free energy to the
Debye-H\"uckel limit at small $\Gamma$, and  $c_\mathrm{TF}$
at large $Z$ is given by the Thomas-Fermi theory
\cite{Salpeter61}.

\subsection{Nonideal mixtures of ions}

A common approximation for the excess (nonideal) free
energy of the strongly coupled ion 
mixture is the LMR,
\beq
f_\mathrm{ex}^\mathrm{LM}(\Gamma) 
\approx \sum_j x_j f_\mathrm{ex}(\Gamma_j,x_j=1)\,,
\label{LMR}
\eeq
where superscript ``LM'' denotes
the linear-mixing approximation,
and all $\Gamma_j$ correspond to the same $\Gamma_e$ 
(assuming that the pressure is given almost totally by the 
strongly degenerate electrons):
$\Gamma_j=\Gamma \,Z_j^{5/3}/\langle Z^{5/3}\rangle$.
In \req{LMR}, $f_\mathrm{ex}$ is the reduced nonideal part of the free
energy: $f_\mathrm{ex}=f_{ii}$ for the ``rigid'' (uniform)
charge-neutralizing electron background and $f_\mathrm{ex} =
f_{ii}+f_{ie}+ Z_j f_{ee}$
for the polarizable background 

The high accuracy of \req{LMR} for binary ionic
mixtures in the rigid background was first demonstrated
by calculations in the HNC approximation \cite{HTV77} and
confirmed later by MC simulations (e.g.,
\cite{DWSC96,Rosenfeld,DWS03}).

\begin{figure}
    \epsfxsize=\columnwidth
    \epsfbox{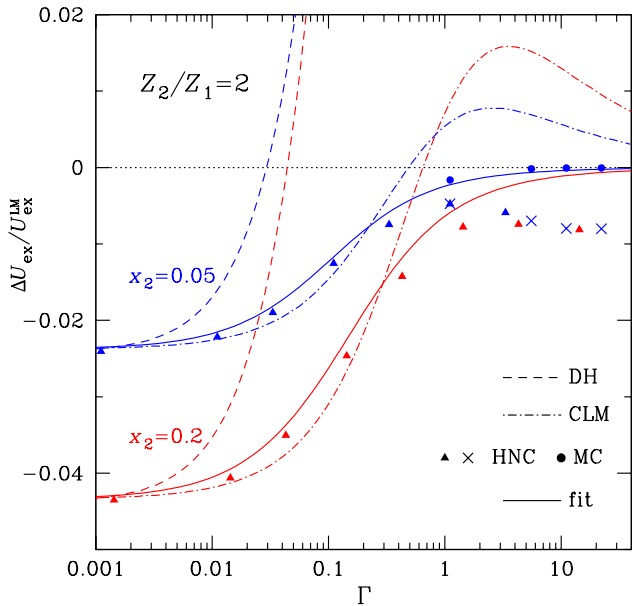}
\caption{
Fractional difference between the Coulomb
part of the internal energy ($U_{ii}$)
in different approximations and the LMR
prediction as a function of the average ion Coulomb
coupling parameter $\Gamma$ for binary mixtures of ions 
with $Z_2/Z_1=2$ and $x_2=1-x_1=0.05$ and 0.2
in the rigid background.
Dashed lines (DH): Debye-H\"uckel formula,
dot-dashed lines (CLM): corrected linear mixing 
\cite{NadyozhinYudin05};
dots: MC results of Ref.~\cite{DWSC96} for $x_2=0.05$;
crosses: HNC results of Ref.~\cite{DWSC96};
triangles: present HNC results;
solid lines: present fit.
}
\label{fig:umix2}
\end{figure}

\begin{figure}
    \epsfxsize=\columnwidth
    \epsfbox{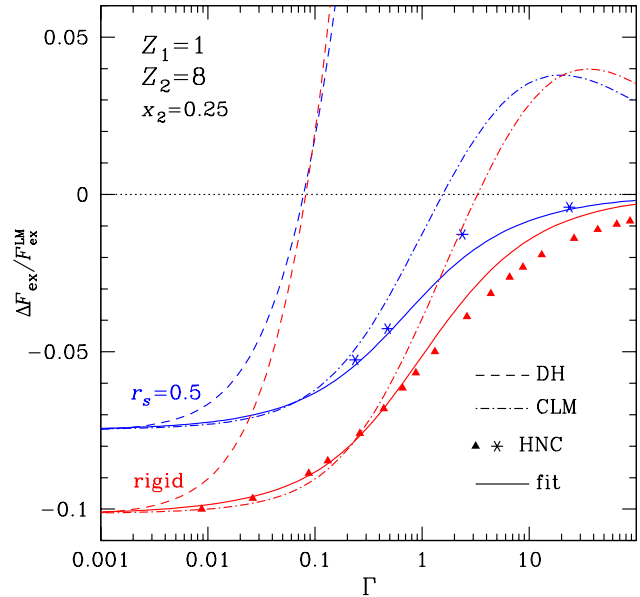}
\caption{
Fractional differences between the nonideal
part of the free energy due to the ion-ion
and ion-electron interactions
($F_{ii}+F_{ie}$) 
in different approximations and the LMR
prediction for a mixture of ions 
with $Z_1=1$ and $Z_2=8$ and $x_2=1-x_1=0.25$
in the rigid and polarizable electron background, as functions of
$\Gamma$.
Dashed lines: DH,
dot-dashed lines: CLM \cite{NadyozhinYudin05};
asterisks: HNC results of Ref.~\cite{ChabAsh};
solid lines: present fit.
}
\label{fig:fmix8}
\end{figure}

\begin{figure}
    \epsfxsize=\columnwidth
    \epsfbox{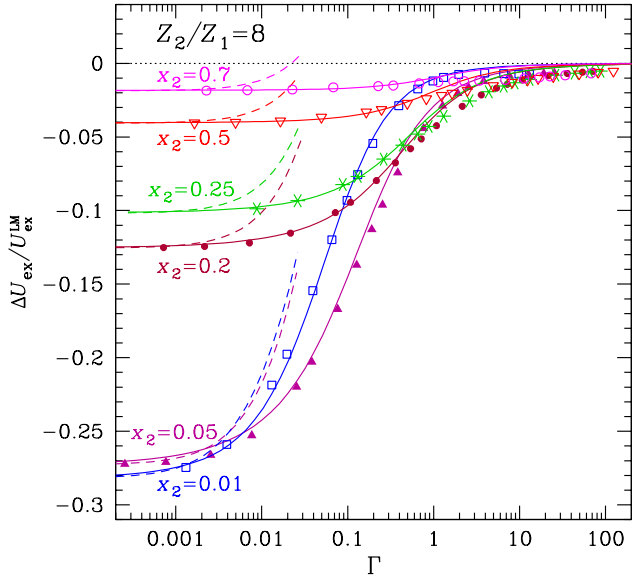}
\caption{
Fractional difference between the Coulomb
part of the internal energy
in different approximations and the LMR
prediction for a binary mixture of ions 
with $Z_2/Z_1=8$ in the rigid background.
Dashed lines (DH): Debye-H\"uckel formula,
symbols: present HNC results;
solid lines: present fit. Different symbols correspond to
different $x_2$ values: 0.01 (squares), 0.05 (solid triangles),
0.2 (dots), 0.25 (asterisks), 0.5 (empty triangles), and 0.7
(empty circles).
}
\label{fig:umix8}
\end{figure}

\begin{figure}
    \epsfxsize=\columnwidth
    \epsfbox{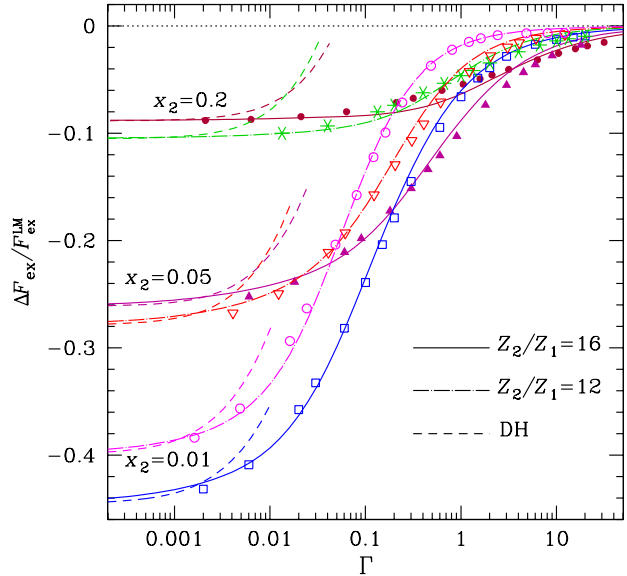}
\caption{
Fractional difference between the Coulomb
part of the free energy
and the LMR
prediction for binary ionic mixtures
with $Z_2/Z_1=12$ and $Z_2/Z_1=16$,
and with $x_2=0.01$, 0.05, and 0.2.
Dashed lines (DH): Debye-H\"uckel formula,
symbols: present HNC results;
solid lines: present fit for $Z_2/Z_1=16$;
long-dash-dot lines: present fit for $Z_2/Z_1=12$.
Different symbols correspond to
different combinations of $x_2$ and $Z_2/Z_1$ values.
}
\label{fig:fmix12_16}
\end{figure}

\begin{figure}
    \epsfxsize=\columnwidth
    \epsfbox{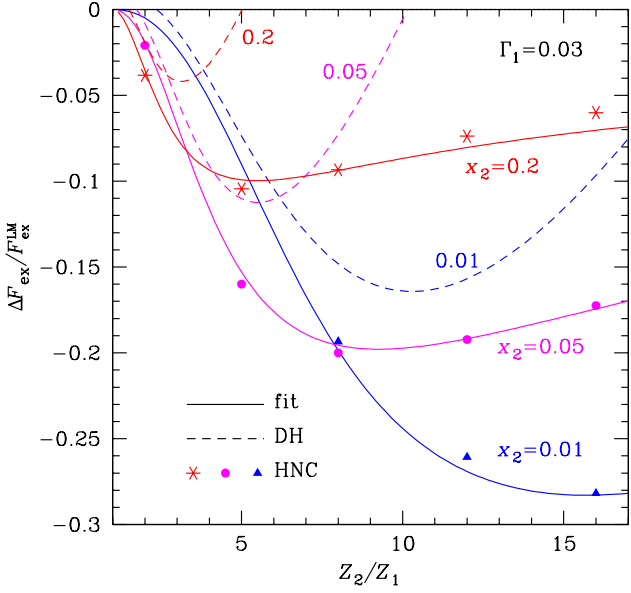}
\caption{
Fractional difference between the Coulomb
part of the internal energy
and the LMR
prediction for binary ionic mixtures
at $\Gamma_1=0.03$ as function of 
the ion charge ratio $Z_2/Z_1$.
Dashed lines (DH): Debye-H\"uckel formula,
symbols: present HNC results;
solid lines: present fit.
Different symbols correspond to
different $x_2$ values:
$x_2=0.01$ (triangles), 0.05 (dots), and 0.2 (asterisks).
}
\label{fig:fmixz03}
\end{figure}

\begin{figure}
    \epsfxsize=\columnwidth
    \epsfbox{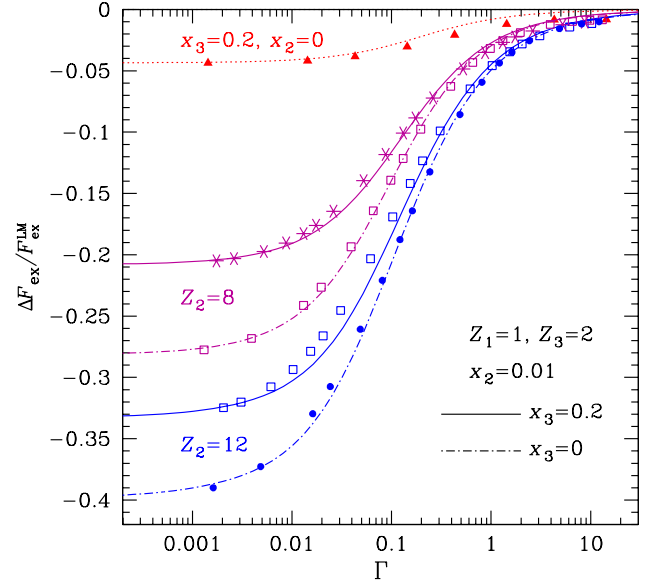}
\caption{
Fractional difference between the Coulomb
part of the free energy
and the LMR
prediction for mixtures of three (solid lines)
and two (dot-dashed and dotted lines) types of ions
with charge ratios $Z_j/Z_1=2$, 8, and 12.
Different symbols correspond to HNC results with
different combinations of $x_j$ and $Z_j$. Solid lines:
$Z_2/Z_1=8$ or 12
($Z_2$ is marked near the curves, assuming $Z_1=1$),
$x_2=0.2$;
$Z_3/Z_1=12$, $x_3=0.01$. Dot-dashed lines: the same 
but without the third kind of ions ($x_3=0$).
Dotted line: binary mixture
with charge ratio $Z_3/Z_1=2$ ($x_2=0$) and number fraction
$x_3=0.2$.
}
\label{fig:fmix3b}
\end{figure}

\begin{figure}
    \epsfxsize=\columnwidth
    \epsfbox{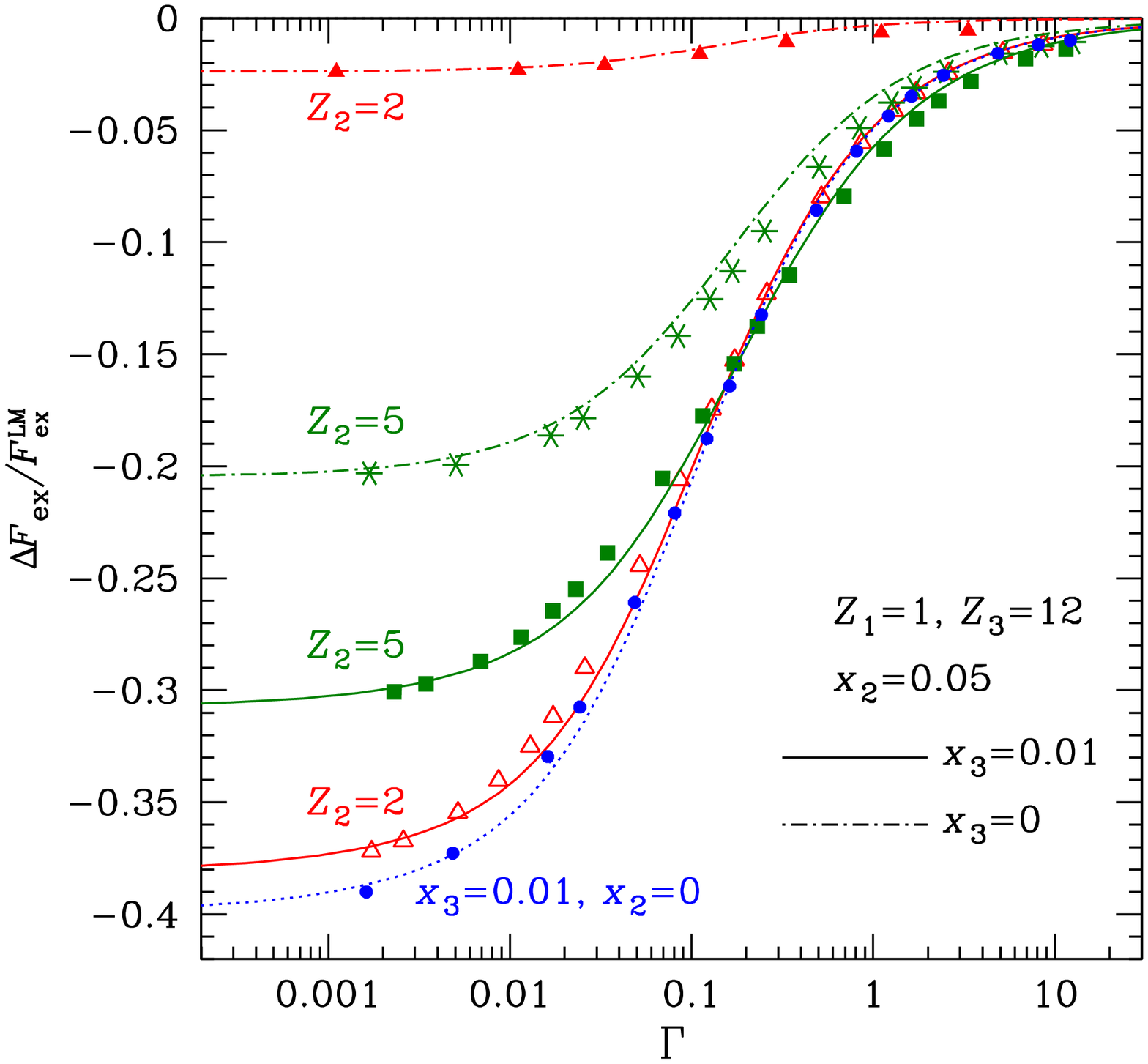}
\caption{
Same as in Fig.~\ref{fig:fmix3b},
but for different charge ratios and
fractional abundances. Solid lines:
$Z_2/Z_1=2$ and 5 (marked near the curves),
$x_2=0.05$; $Z_3/Z_1=12$, $x_3=0.01$. Dot-dashed lines: the same 
but with $x_3=0$. Dotted line: the same with $x_2=0$
($x_3=0.01$).
}
\label{fig:fmix3c}
\end{figure}

The validity of the LMR in the case of an ionic mixture immersed 
in a \emph{polarizable} finite-temperature electron background
has been examined by Hansen et al.~\cite{HTV77} in the 
first-order thermodynamic perturbation approximation and  by
Chabrier and Ashcroft \cite{ChabAsh} by solving the HNC equations
with effective screened potentials. These authors found that the
LMR remains accurate when the electron
response is taken into account in the inter-ionic potential, as
long as the Coulomb coupling is strong ($\Gamma>1$).

On the other hand, the LMR is invalid at $\Gamma\ll1$.
Indeed, in this case the Debye-H\"uckel theory gives
\beq
   f_{ee}^\mathrm{DH} = - \Gamma_e^{3/2}/\sqrt{3},
\quad
   f_{ii}^\mathrm{DH} = f_{ee}^\mathrm{DH}
   \zeta_{ii}^\mathrm{DH},
\eeq
and for the EIP
\beq
   f_\mathrm{ex}^\mathrm{DH}
= f_{ii}^\mathrm{DH}+f_{ie}^\mathrm{DH}+ \langle Z \rangle f_{ee}^\mathrm{DH}
   = f_{ee}^\mathrm{DH}
   \zeta_\mathrm{eip}^\mathrm{DH},
\eeq
where
\beq
   \zeta_{ii}^\mathrm{DH} =
    \frac{\langle Z^2\rangle^{3/2}}{\langle Z\rangle^{1/2}},
\quad   
      \zeta_\mathrm{eip}^\mathrm{DH} =
   \,\frac{(\langle Z^2\rangle+\langle Z\rangle)^{3/2}
   }{\langle Z\rangle^{1/2}} .
\eeq
However, the LMR at $\Gamma\ll1$ gives another result:
\beq
   f_{ii}^\mathrm{LM} \sim 
   f_{ee}^\mathrm{DH} \zeta_{ii}^\mathrm{LM},
\quad
   f_\mathrm{ex}^\mathrm{LM}
   \sim  
   f_{ee}^\mathrm{DH} \zeta_{eip}^\mathrm{LM},
\eeq
where
\beq
   \zeta_{ii}^\mathrm{LM} = 
   \langle Z^{5/2}\rangle,
\quad
   \zeta_\mathrm{eip}^\mathrm{LM} = 
   \langle Z\,(Z+1)^{3/2}\rangle .
\eeq

Nadyozhin and Yudin \cite{NadyozhinYudin05} considered several
possible modifications of the LMR at intermediate
$\Gamma$ and found that such modification can
appreciably shift the statistical nuclear equilibrium at the
conditions typical of the final stage of a stellar gravitational
collapse. They considered the rigid 
background and advocated a modification of every
term in \req{LMR} by multiplying the leading
fit coefficient at small $\Gamma$ by factor
$d_j = \sqrt{\langle Z^2\rangle/Z_j\langle Z\rangle}$.
It corresponds to replacing $\sqrt{3}/2$ in \req{A3}
by $d_j\,\sqrt{3}/2$ (and a possible simultaneous change of
$A_2$). For a compressible background an analogous modification
implies additionally replacement of \req{cDH} by
\beq
   c_\mathrm{DH}^{(j)} = \frac{Z_j}{\sqrt{3}}
   \,\left[\left(\frac{\langle Z^2\rangle}{\langle Z\rangle} + 1
   \right)^{3/2}
   -\frac{\langle Z^2\rangle^{3/2}}{\langle Z\rangle^{3/2}} - 1
   \right].
\label{CDH_CLM}
\eeq
Hereafter modifications of this type will be
called \emph{corrected linear mixing} (CLM).

The result of such modifications is shown by dot-dashed lines in
Figs.~\ref{fig:umix2} and \ref{fig:fmix8}, where we plot the
ratios $\Delta u_{ii}/u_{ii}^\mathrm{LM}$
and $\Delta f_\mathrm{ex}/f_\mathrm{ex}^\mathrm{LM}$
as functions of $\Gamma$. Here  $\Delta
f\equiv f-f^\mathrm{LM}$  and $\Delta u\equiv u-u^\mathrm{LM}$
are the deviations of the reduced free and internal energies,
respectively, from the LMR. In Fig.~\ref{fig:umix2} the
electron response  is neglected (rigid background). We see that,
for example, at $\Gamma\approx10$
the CLM prescription gives $\Delta u$ and
$\Delta f$ corrections of about 1\% in Fig.~\ref{fig:umix2} and
several percents in Fig.~\ref{fig:fmix8}, whereas
they must be much smaller according to the MC results
\cite{DWSC96,DWS03}. Moreover, $\Delta u$ and $\Delta f$ in the CLM
approximation have the incorrect sign at $\Gamma\gtrsim1$ (note
that according to Ref.~\cite{HTV77} $\Delta u/u$ and $\Delta f/f$
are negative at any $\Gamma$). Additional modifications of 
the coefficient $A_2$ in
\req{LMR} also do not solve the problem.

An alternative to the CLM, named ``complex mixing''
in Ref.~\cite{NadyozhinYudin05}, maintains the correct sign
of $\Delta f$ and $\Delta u$, but leads to still larger absolute
values of these corrections (i.e., still slower recovery of the
LMR) at large $\Gamma$.

In order to find a more accurate approximation, we have performed
HNC calculations of $F_{ii}$, $U_{ii}$ and $P_{ii}$ for binary
ionic mixtures in the rigid background for a
broad $\Gamma$ range and various values of the charge number
ratios $Z_2/Z_1$ and fractional abundances $x_2=1-x_1$. Some of
the results are shown by triangles in Figs.~\ref{fig:umix2} and
\ref{fig:fmix8}. In agreement with the previous studies 
(e.g., \cite{HTV77}), our numerical
results show monotonically decreasing fractional deviations from
the LMR with increasing $\Gamma$. The results agree to at least
5 digits with those published in \cite{DWSC96} (crosses in
Fig.~\ref{fig:umix2}). At $\Gamma\gtrsim1$, HNC results tend to a
constant residual within 1\%, which is due to the intrinsic
inaccuracy of the HNC approximation for strongly coupled plasmas
because of the lack of the bridge functions in the diagrammatic
representation of this approximation. To prove this statement,
in Fig.~\ref{fig:umix2} we plot by dots the values of $\Delta
u_{ii}/u_{ii}^\mathrm{LM}$ from MC simulations 
\cite{DWSC96}. The latter simulations give tiny deviations from
the LMR at $\Gamma\gtrsim3$, which are invisible in the figure
scale.

Asterisks in Fig.~\ref{fig:fmix8} correspond to the HNC
calculations for mixtures in the \emph{polarizable} electron
background \cite{ChabAsh}, which give qualitatively the same
results as the calculations for the rigid background. The
second dot-dashed curve in this figure shows CLM for the
polarizable background, according to \req{CDH_CLM}.

A correction to the linear mixing rule, which
exactly recovers the Debye limit at $\Gamma\to0$ and the LMR at
$\Gamma\gg1$, and which agrees with the HNC data, can be
expressed by the following analytic fitting formula:
\beq
   \Delta f = f_{ee}^\mathrm{DH}
 \,\frac{\zeta^\mathrm{DH}-\zeta^\mathrm{LM}}{
 (1 + a \Gamma^{b})^{c}},
\label{fit}
\eeq
where $\zeta=\zeta_{ii}$ or $\zeta_\mathrm{eip}$ for the rigid
or polarizable background, respectively, and
the parameters $a$, $b$, and $c$ depend on plasma
composition as follows:
\bea
   a &=& \frac{2.2\,\delta+17\,\delta^4}{1-b},
\quad
   \delta = \frac{\zeta^\mathrm{LM}-\zeta^\mathrm{DH}}{\langle
   Z^{5/2} \rangle} ,
\nonumber\\
   b &=& d^{-0.2}, 
\quad
   c = 1+d/6 ,
\quad
   d = \langle Z^2 \rangle / \langle Z \rangle^2.
\nonumber
\eea
Thanks to the simple form of this formula, its derivatives are also 
rather simple. For example, the corrections to the reduced
internal energy and heat capacity read, respectively,
\bea
   \Delta u &=& \left( \frac32
    - \frac{abc\,\Gamma^b}{1 + a \Gamma^{b}}\right) \Delta f,
\\
   \Delta c &=& \left(
    \frac{abc\,\Gamma^b}{1 + a \Gamma^{b}} - \frac12 \right)
    \Delta u
     -\frac{ab^2c\,\Gamma^b}{(1 + a \Gamma^{b})^2}
     \Delta f.
\eea

This approximation has been compared with our HNC calculations
for binary ionic mixtures in the rigid electron background at
$Z_2/Z_1=2$, 5, 8, 12, and 16, with $x_2=1-x_1$ ranging from 0.01
to 0.7. Some of the results are shown in the figures.  In all
figures the Debye-H\"uckel approximation is drawn by dashed lines
for comparison. In Figs.~\ref{fig:umix2} and \ref{fig:fmix8},
discussed above, and in Figs.~\ref{fig:umix8},
\ref{fig:fmixz03} our approximation is drawn by solid lines.
Figure~\ref{fig:umix8} shows the fractional difference between the
different results for internal energy and the LMR approximation
for binary ionic mixtures with $Z_2/Z_1=8$ and $x_2=0.01$, 0.05,
0.2, 0.25, 0.5, and 0.7 (marked near the curves). In
Fig.~\ref{fig:fmix12_16} we show the case of the
highest considered
charge asymmetries: long-dash-dot lines show the
approximation (\ref{fit}) for $Z_2/Z_1=12$ and solid lines for
$Z_2/Z_1=16$. Finally, in Fig.~\ref{fig:fmixz03} we show a
dependence of $\Delta f_{ii}/f_{ii}^\mathrm{LM}$ on the charge ratio
$Z_2/Z_1$ at $\Gamma_1=0.03$. Here
different symbols correspond to the HNC
results for different values of $x_2$.

The
difference between the HNC results 
and formula (\ref{fit}) for $f_\mathrm{ex}$
lies within 0.013 and within 1\%. The internal
energy calculated using the analytic derivative of the fit
(\ref{fit}) deviates from the HNC results by not more than
0.017 and not more than 1.5\%. These maximal deviations are attained for the extremely
asymmetric mixtures with $Z_2/Z_1=16$. 

The analytic formula (\ref{fit}) has also been
compared to available HNC results for binary ionic
mixtures in the polarizable electron background \cite{ChabAsh}
and found to be satisfactory within the accuracy of the latter
results (an example is shown in Fig.~\ref{fig:fmix8}).

We have also performed calculations for mixtures of ions
of three different types on the rigid electron
background and compared the results
with \req{fit}. The results
of the comparison are shown in Figs.~\ref{fig:fmix3b}
and \ref{fig:fmix3c}. Solid lines show the difference of
the Coulomb free energy from the LMR according to \req{fit}
for 3-component mixtures; for comparison, dot-dashed and dotted lines 
are plotted for 2-component mixtures with the same charge ratios;
symbols represent the HNC results.
In all considered cases, adding a third component to a binary mixture
increases the deviations of \req{fit} from HNC results
by less than a factor of 1.5.
We conclude that the agreement between the fit and numerical
results remains satisfactory.

\section{Conclusions}
\label{sect:concl}

We have performed HNC calculations of
the free energy, internal energy, and pressure for various
ionic mixtures with different fractional abundances of the ion
species in a broad range of $Z$ and $\Gamma$ values. We have
constructed an analytic approximation to the deviation from the
LMR, which recovers the Debye-H\"uckel formula for multicomponent
plasmas at $\Gamma\ll1$ and the LMR at $\Gamma\gg1$, and which
describes our calculations at any $\Gamma$ values, as well as 
the HNC and MC results for the internal
energy of plasma mixtures, available
in the literature, with an accuracy better than 2\%.\\

\begin{acknowledgments}
The work of G.C.\ and A.Y.P.\ was partially supported
by the CNRS French-Russian Grant No.\ PICS 3202.
The work of A.Y.P.\ was partially supported 
by the Rosnauka Grant NSh-2600.2008.2
and the RFBR Grant 08-02-00837. The work of F.J.R.\ was
partially performed under the auspices of the U.S. Department of
Energy by Lawrence Livermore National Laboratory under Contract
DE-AC52-07NA27344.
\end{acknowledgments}


\end{document}